\documentclass{article}
\usepackage{ijcai25}

\usepackage{times}
\usepackage{soul}
\usepackage{url}
\usepackage[hidelinks]{hyperref}
\usepackage[utf8]{inputenc}
\usepackage[small]{caption}
\usepackage{graphicx}
\usepackage{amsmath}
\usepackage{amsthm}
\usepackage{booktabs}
\usepackage{algorithm}
\usepackage{algorithmic}
\usepackage[switch]{lineno}
\usepackage{multirow}

\usepackage{amssymb}
\usepackage{subcaption} 

\usepackage{comment}
\usepackage{pifont}
\usepackage{changepage} 
\usepackage{lipsum} 
\usepackage{wrapfig} 
\usepackage{mathrsfs}
\usepackage{tabularx}
\usepackage{authblk}
\DeclareUnicodeCharacter{0308}{\"}
\newtheorem{definition}{Definition}
\newtheorem{proposition}{Proposition}

\title{PDDFormer: Pairwise Distance Distribution Graph Transformer\\
for Crystal Material Property Prediction}

\author[1]{Xiangxiang Shen\textsuperscript{*}}
\author[2]{Zheng Wan\textsuperscript{*}}
\author[1]{Lingfeng Wen}
\author[1]{Licheng Sun}
\author[3]{Jian Yang}
\author[4]{Xuan Tang}
\author[5]{Shing-Ho J. Lin}
\author[2]{Xiao He}
\author[1]{Mingsong Chen}
\author[1]{Xian Wei\textsuperscript{†}}
\affil[1]{Software Engineering Institute, East China Normal University}
\affil[2]{School of Chemistry and Molecular Engineering, East China Normal University}
\affil[3]{School of Geospatial Information, Information Engineering University}
\affil[4]{School of Communication and Electronic Engineering, East China Normal University}
\affil[5]{School of Artificial Intelligence, University of Chinese Academy of Sciences}

\begin{document}
\maketitle

\begin{abstract}
Crystal structures can be simplified as a periodic point set that repeats across three-dimensional space along an underlying lattice. 
Traditionally, crystal representation methods characterize the structure using descriptors such as lattice parameters, symmetry, and space groups. 
However, in reality, atoms in materials always vibrate above absolute zero, causing their positions to fluctuate continuously. 
This dynamic behavior disrupts the fundamental periodicity of the lattice, making crystal graphs based on static lattice parameters and conventional descriptors discontinuous under slight perturbations. 
Chemists proposed the pairwise distance distribution (PDD) method to address this problem. However, the completeness of PDD requires defining a large number of neighboring atoms, leading to high computational costs. Additionally, PDD does not account for atomic information, making it challenging to apply it directly to crystal material property prediction tasks. 
To tackle these challenges, we introduce the atom-Weighted Pairwise Distance Distribution (WPDD) and Unit cell Pairwise Distance Distribution (UPDD) and apply them to the construction of multi-edge crystal graphs. 
We demonstrate the continuity and general completeness of crystal graphs under slight atomic position perturbations. 
Moreover, by modeling PDD as global information and integrating it into matrix-based message passing, we significantly reduce computational costs. 
Comprehensive evaluation results show that WPDDFormer achieves state-of-the-art predictive accuracy across tasks on benchmark datasets such as the Materials Project and JARVIS-DFT.
\end{abstract}
\section{Introduction}
\begin{figure*}[t]
\centering
\includegraphics[width=0.91\textwidth]{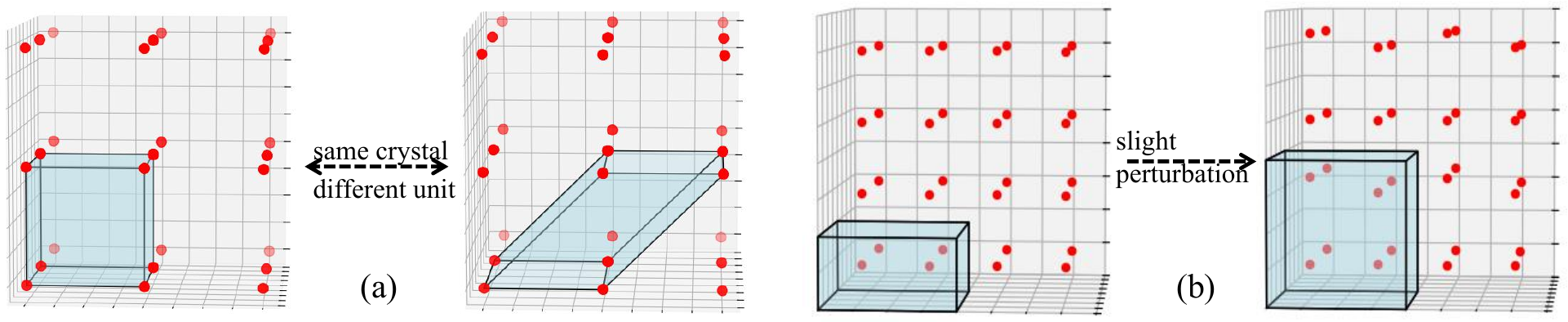}
\caption{Illustrations of different unit cells and lattice representations of the same crystal structure. Figure (a) shows several possible choices among the infinitely many unit cells for the same crystal structure in the undisturbed case. Figure (b) illustrates that for almost any perturbation, the symmetry group and any reduced unit cell (with minimal volume) will undergo discontinuous changes.}
\label{lattice}
\end{figure*}
Crystals are solids with a regular geometric shape formed by atoms, ions, or molecules arranged periodically in space during the crystallization process.
Their structure is typically described using repeating unit cells and lattice vectors. However, this description method brings a fundamental challenge: the same crystal structure can be represented by different unit cells and lattice vectors, as shown in Figure \ref{lattice}(a). 
Additionally, in real-world scenarios, the experimental coordinates of unit cells and atoms are inevitably affected by atomic vibrations and measurement noise. 
These subtle disturbances can lead to discontinuous changes in any simplified unit cell~\cite{kurlin2024mathematics}, resulting in numerous unit cells for a given crystal structure, as shown in Figure \ref{lattice}(b), thereby introducing ambiguity in the representation of crystal data~\cite{widdowson2022resolving}.
Currently, many graph neural networks~\cite{batzner20223,yan2022periodic,yan2024complete,yan2024space} typically use unit cell parameters, simplified cell parameters, symmetry, and space groups to represent the periodic structure of crystals. 
However, these features are either non-invariant or discontinuous~\cite{zwart2008surprises} invariants, leaving the issue of ambiguity in crystal data unresolved~\cite{patterson1944ambiguities,widdowson2022average,groom2016cambridge,bartok2013representing,wassermann2010activity,ahmad2018machine}. 

The continuous and complete invariant—Pairwise Distance Distribution (PDD)—proposed by~\cite{widdowson2022resolving} addresses the ambiguity in crystal data representation by distinguishing all periodic structures in the world’s largest real material collection, the Cambridge Structural Database. 
To achieve completeness, PDD requires a predetermined number of sufficient neighbors, which is computationally expensive and difficult to directly apply for predicting crystal properties~\cite{balasingham2022compact}. 
~\cite{balasingham2024material} employed distance distribution graphs (DDGs) based on PDD to predict the properties of crystal materials, but they did not achieve satisfactory performance (only slightly better than CGCNN), and although this approach reduced computational costs, it compromised the completeness of PDD. 
In contrast, crystal graph representations based on multi-edge crystal graphs and unit cell parameters~\cite{yan2024complete} achieve completeness, more accurately characterizing crystal structures, and achieving state-of-the-art performance in crystal material property prediction tasks. 
However, using unit cell parameters leads to discontinuities in the crystal graphs.

Since PDD does not account for atomic types, it is challenging to use it directly for effective crystal property prediction. To better represent crystal structures, we first introduce WPDD and UPDD.
Furthermore, we integrate WPDD and UPDD into the construction of multi-edge crystal graphs and propose the PDD Graph Transformer (including WPDDFormer and UPDDFormer) based on the transformer architecture. 
We model WPDD as global information and incorporate it into matrix-based message passing without introducing additional information to capture periodicity, significantly reducing computational costs (as shown in Table \ref{efficency}). 
Finally, we employ the Earth Mover's Distance (EMD)~\cite{rubner2000earth} to assess the continuity of crystal graphs, demonstrating that WPDD crystal graphs constructed using only Euclidean distances maintain continuity and general completeness under slight atomic position perturbations, providing a more accurate depiction of actual crystal structures. 
Ablation experiments show the crucial role of (W/U)PDD in constructing crystal graphs. Through comprehensive evaluations, our method achieves state-of-the-art predictive accuracy across various tasks in the Materials Project and JARVIS datasets. 
This advancement highlights the effectiveness of WPDDFormer in bridging the gap between traditional crystal descriptors and dynamic atomic behavior, leading to more accurate and reliable predictions in materials science.

\section{Preliminaries}

\subsection{The Structure of Crystals}
\label{crystal}
By selecting an appropriate structural unit, the entire crystal structure can be viewed as the periodic repetition of this unit in space. 
This property, where atoms within a crystal repeat in three-dimensional space according to a specific pattern, is called periodicity, with the smallest repeatable structural unit being the unit cell. 
The unit cell can be defined as $\mathcal{U}= \left ( \mathcal{X},\mathcal{P}  \right ) $, where $\mathcal{X}$ and $\mathcal{P}$ can be represented in matrix form. 
Typically, $\mathcal{X} = \left [ x_{1},x_{2}  \cdots x_{n-1},x_{n}\right ]^{T}\in \mathbb {R} ^{n\times 1}$, where 
$n$ represents the number of atoms and $ {x_{i}}\in \mathbb{R} ^{1}$ represents the atomic type of atom $ {i}$ in the unit cell. 
$\mathcal{P} = \left [ p_{1},p_{2}  \cdots p_{n-1},p_{n}\right ]^{T}\in \mathbb {R} ^{n\times 3}$ is the atomic position matrix, where $ {p_{i}}\in \mathbb {R} ^{3}$ represents the Cartesian coordinates of the atom $ {i}$ in the unit cell in 3D space. 
The lattice vectors $\mathcal{L} = \left [ l_{1},l_{2},l_{3}\right ]^{T}\in \mathbb {R}^{3\times3}$ can reflect the way the unit cell repeats in three directions to map the periodic crystal structure.
Therefore, in 3D space, the infinite crystal structure $\mathcal{S}$ can be represented as $\left ( \mathcal{U},\mathcal{L}  \right )$.

\subsection{Definitions}
\begin{definition}
\textbf{Pointwise Distance Distribution.} 
    For the infinite crystal structure $\mathcal{S}=\left ( \mathcal{U},\mathcal{L}  \right )$ mentioned in Section \ref{crystal}, fix a neighbor count $k\ge1$. For each point $x_{i}$ in the unit cell $\mathcal{U}$, let $d_{i1} \le \dots \le d_{ik}$ be the Euclidean distances from $\textbf{p}_{i}$ to its $k$ nearest neighbors in the infinite crystal structure. 
Consider an $n \times k$ matrix composed of $n$ rows of distance vectors, where each point $x_{i} \in \mathcal{U}$ corresponds to one row. 
If the matrix contains $m\ge1$ identical rows, they are merged into one row with a weight of $\frac{m}{n}$. The resulting matrix can be regarded as a weighted distribution of rows, which is called the Pairwise Distance Distribution $\mathcal{PDD}$$\left ( S;k \right )\in \mathbb {R} ^{n\times (k+1)}$.
\end{definition}

According to~\cite{widdowson2022resolving} and~\cite{yan2024complete}, we present Definitions 2-3. According to~\cite{widdowson2022resolving}, we present Definitions 4-6.

\begin{definition}
\textbf{Isometric Crystal Graphs.} 
An isometric transformation is a mapping that preserves Euclidean distances, denoted as $f\left ( x \right ) =Rx+b$. Any isometric transformation $f$ can be decomposed into translation, rotation, and reflection.
Specifically, suppose there exists a rotation matrix $ R\in \mathbb {R}^{3\times3 }$, with a determinant of 1 ($\left | R \right | =1$), and a translation vector $b\in \mathbb {R}^{3} $, then two crystal structures $\mathcal{S}=\left ( \mathcal{U},\mathcal{L}  \right )$ and $\mathcal{Q}=\left ( \mathcal{U}' ,\mathcal{L}'   \right )$ are isometric, satisfying $\mathcal{U}'=R\mathcal{U}+b$, where $R\mathcal{U}+b$ denotes the application of the rotation $R$ and translation $b$ to each element in the infinite set $\mathcal{U}$.
\end{definition}
Suppose $\mathcal{S}$ and $\mathcal{Q}$ are isometric. In that case, their crystal graph representations satisfy $\mathcal{G}\left ( \mathcal{S}  \right ) =\mathcal{G}\left ( \mathcal{Q}  \right )$, which means that the graphical representation of the crystal structure produces no false positives; that is, there are no isometric pairs where $\mathcal{G}\left ( \mathcal{S}  \right ) \neq \mathcal{G}\left ( \mathcal{Q}  \right )$ but $\mathcal{S} \simeq \mathcal{Q}$. Conversely, if $\mathcal{G}\left ( \mathcal{S}  \right ) =\mathcal{G}\left ( \mathcal{Q}  \right )$, then $\mathcal{S}$ and $\mathcal{Q}$ are isometric, meaning f produces no false negatives, i.e., there are no non-isometric pairs where $\mathcal{G}\left ( \mathcal{S}  \right ) =\mathcal{G}\left ( \mathcal{Q}  \right )$ but $\mathcal{S} \not\simeq \mathcal{Q}$.
That is, if the crystal graph representations of artificially constructed crystal structures are identical under isometric transformations, then they are geometrically equivalent.

\begin{definition}
\textbf{Geometrically Complete Crystal Graphs.}  
If we construct crystal graphs $\mathcal{G}\left ( \mathcal{S}  \right ) = \mathcal{G}\left ( \mathcal{Q}  \right )\Longrightarrow \mathcal{S} \simeq \mathcal{Q}$, where $\simeq$ denotes the isomorphism of two crystals as defined in Definition 2, then the crystal graph $\mathcal{G}$ is geometrically complete. This means that if two crystal graphs $\mathcal{G}\left ( \mathcal{S}  \right )$ and $\mathcal{G}\left ( \mathcal{Q}  \right )$ are identical, the infinite crystal structures represented by $\mathcal{G}\left ( \mathcal{S}  \right )$ and $\mathcal{G}\left ( \mathcal{Q}  \right )$ are also identical. 
If the constructed crystal graph $\mathcal{G}$ can distinguish any subtle structural differences between different crystal materials, it is said to be geometrically complete.
\end{definition}

\begin{definition}
\textbf{Metric.}  
The metric $d$ between crystal graphs $\mathcal{G}$ satisfies all the axioms: 1) $d\left ( \mathcal{G}\left ( \mathcal{S}  \right ) =\mathcal{G}\left ( \mathcal{Q}  \right ) \right )=0 $ if and only if $\mathcal{G}\left ( \mathcal{S}  \right ) =\mathcal{G}\left ( \mathcal{Q}  \right )$; 2) Symmetry: $d\left ( \mathcal{G}\left ( \mathcal{S}  \right ) , \mathcal{G}\left ( \mathcal{Q}  \right ) \right ) =d\left ( \mathcal{G}\left ( \mathcal{Q}  \right ) ,\mathcal{G}\left ( \mathcal{S}  \right ) \right ) $; 3) Triangle inequality: $d\left ( \mathcal{G}\left ( \mathcal{S}  \right ), \mathcal{G}\left ( \mathcal{Q}  \right ) \right ) + d\left ( \mathcal{G}\left ( \mathcal{Q}  \right ), \mathcal{G}\left ( \mathcal{K}  \right ) \right ) \ge   d\left ( \mathcal{G}\left ( \mathcal{S}  \right ), \mathcal{G}\left ( \mathcal{K}  \right ) \right ) $.
\end{definition}

\begin{definition}
\textbf{Lipschitz continuity of crystal graphs.}  
If $\mathcal{Q}$ is obtained by moving each point in the periodic crystal $\mathcal{S} \subset \mathbb{R}^{n}$ by no more than $\epsilon$, and the distance of the constructed crystal graph structures satisfies $d\left ( \mathcal{G}\left ( \mathcal{S}  \right ), \mathcal{G}\left ( \mathcal{Q}  \right ) \right )\leq C \epsilon$, where $C$ is a constant, then the crystal graph is continuous, and $\mathcal{Q},\mathcal{S} \subset \mathbb{R}^{n}$ can be any periodic crystal structures.
\end{definition}

\begin{definition}
\textbf{EMD.}  
Let $\mathcal{G} \left ( \mathcal{S}  \right )$ and $\mathcal{G} \left ( \mathcal{Q}  \right )$ be the crystal graph structures we construct for periodic crystals $\mathcal{S}$ and $\mathcal{Q}\in \mathbb{R}^{n}$ . The flow from $\mathcal{G} \left ( \mathcal{S}  \right )$ to $\mathcal{G} \left ( \mathcal{Q}  \right )$ is represented by an $n \left ( \mathcal{S}  \right )\times n \left ( \mathcal{Q} \right )$ matrix, where the elements $f_{ij} \in [0, 1]$ indicate the partial flow from $\mathcal{R}_{i} \left ( \mathcal{S}  \right )$ to $\mathcal{R}_{j} \left ( \mathcal{Q}  \right )$. The Earth Mover's Distance (EMD) is defined as the minimum cost:
\begin{equation}
\begin{split}
\resizebox{.87\linewidth}{!}{$\mathrm{EMD}\left ( \mathcal{G} \left ( \mathcal{S}  \right ), \mathcal{G} \left ( \mathcal{Q}  \right ) \right )=\sum_{i=1}^{n}\sum_{j=1}^{n}f_{ij}\left | R_{i} \left ( \mathcal{S}  \right ) -R_{j} \left ( \mathcal{Q}  \right ) \right |$}
\\
\resizebox{.91\linewidth}{!}{$s.t. \sum_{i=0}^{n} f_{ij} \le w_{i} \left ( \mathcal{S}  \right ) , \: \sum_{j=0}^{n} f_{ij} \le w_{j} \left ( \mathcal{Q}  \right ) ,\:
\sum_{i=1}^{n}\sum_{j=1}^{n}f_{ij} =1$}
\end{split}
\end{equation}
The first condition $\sum_{i=0}^{n} f_{ij} \le w_{i} \left ( \mathcal{S}  \right )$ means that not more than the weight $w_{i} \left ( \mathcal{S}  \right )$ of the component 
$R_{i} \left ( \mathcal{S}  \right )$ ‘flows’ into all components $R_{j} \left ( \mathcal{Q}  \right )$ via ‘flows’$f_{ij}$ . Similarly, the second condition $\sum_{j=0}^{n} f_{ij} \le w_{j} \left ( \mathcal{Q}  \right )$ means that all ‘flows’ $f_{ij}$ from $R_{i} \left ( \mathcal{S}  \right )$ ‘flow’ 
Into $R_{j} \left ( \mathcal{Q}  \right )$ up to the maximum weight $w_{j} \left ( \mathcal{Q}  \right )$. The last condition $\sum_{i=1}^{n}\sum_{j=1}^{n}f_{ij} =1$ forces to 
‘flow’ all rows $R_{i} \left ( \mathcal{S}  \right )$ to all rows $R_{j} \left ( \mathcal{Q}  \right )$.
\end{definition}

\section{Related Work}
\paragraph{Finite Crystal Graph Representation}
CGCNN~\cite{xie2018crystal} represents crystal structures as finite multi-edge crystal graphs to model crystal structures and predict material properties. Building on the construction of multi-edge crystal graphs, MegNet~\cite{chen2019graph} introduced global state attributes into graph networks, while GATGNN~\cite{louis2020graph} utilized multiple graph attention layers (GAT) to learn the properties of local neighborhoods and employed global attention layers to weight global atomic features. ALIGNN~\cite{choudhary2021atomistic} and M3GNet~\cite{chen2022universal} incorporated angular information into the message-passing process to generate richer and more discriminative representations. ~\cite{das2022crysxpp} proposed an interpretable deep property predictor called CrysXPP. CrysMMNet~\cite{das2023crysmmnet} adopted a multimodal framework, integrating graph and text representations to produce joint multimodal representations of crystalline materials. ~\cite{das2023crysgnn} proposed CrysGNN, a pretraining framework leveraging unlabeled data, while CrysDiff~\cite{song2024diffusion} introduced a diffusion model-based pretraining–fine-tuning framework. However, the methods above represent crystals as finite graph structures, which fail to effectively capture the periodicity of infinite crystals.

\paragraph{Periodic Representation of Crystals}
Recently, Matformer~\cite{yan2022periodic} encoded periodic patterns by adding self-connecting edges to atoms based on lattice parameters. PotNet~\cite{lin2023efficient} considered the infinite summation of interatomic interactions. Crystalformer~\cite{taniai2024crystalformer} performed infinite summations of interatomic potentials through infinitely connected attention while also utilizing lattice parameters. ComFormer~\cite{yan2024complete} constructed cell parameters by adding self-connecting edges to atoms and their copies in three different directions to encode periodic patterns, employing equivariant vector representations and invariant geometric descriptors of Euclidean distances and angles to represent the geometric information of crystals. GMTNet~\cite{yan2024space} aims to predict the tensor properties of crystalline materials while satisfying O(3) group equivariance and the symmetry of crystal space groups. However, the crystal structures they represent rely on non-invariant or discontinuous invariants, failing to resolve the issue of crystal data fuzziness.

\paragraph{Continuity and Complete Representations for Crystals}
Addressing the continuity and completeness of crystal representations is a critical issue. Recent advancements in AMD~\cite{wang2022approximately} and PDD~\cite{widdowson2022resolving} have developed matrix forms that are both complete and continuous. However, in practical applications, using these matrix representations as inputs for predicting crystal properties without compromising continuity and completeness is challenging. The AMD and PDD representations are designed to distinguish stable crystal structures and do not consider atomic types; their completeness assumption only holds for stable structures. Additionally, to achieve completeness, a sufficiently large number of neighbors $k$ must be predetermined for any test crystal. 
Directly modeling PDD as edge information is impractical and costly in real-world applications~\cite{balasingham2022compact}.

\begin{figure*}[t]
\centering
\includegraphics[width=0.89\textwidth]{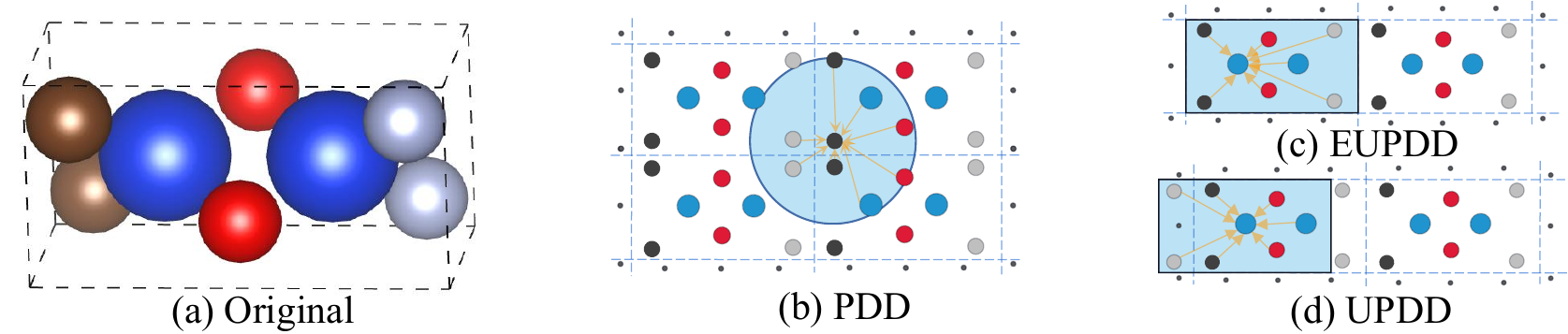} 
\caption{  Schematic diagram of the selected neighbors in PDD. Figure (a) represents the 3D unit cell structure. The edges in Figure (b) show the neighbor selection for atom $i$ in WPDD. By comparing Figures (c) and (d), we can see that we construct the unit cell centered around each atom and select neighbors, rather than being limited to the unit cell where the atoms are located.}
\label{3DPDD}
\end{figure*}

\section{PDDFormer}
In this section, we propose two PDD variants, namely WPDD and UPDD, and then incorporate them into crystal graph construction. We finally present the PDDformer framework.
\subsection{Atom-Weighted PDD (WPDD)}
Since the PDD representation is designed for stable crystal structures and does not consider atomic types, it is not suitable for predicting crystal material properties. 
To account for the influence of atomic types, for a given crystal structure $\mathcal{S} = \mathcal{U} + \mathcal{L}$, where each atom $x_{i} \in \mathcal{U}$ is labeled with the atomic mass $t(x_{i})$ corresponding to it, the final weight for each row is $\mathcal{W} = [ w_{1}, \dots, w_{n} ]^{T}$, where $w_{i} = \frac{t(x_{i})}{ \sum_{u=1}^{n} t(x_{u}) }$. 
By concatenating $\mathcal{W}$ with $\mathcal{PDD} \in \mathbb {R}^{n\times  k}$, an atomic-mass-weighted $\mathcal{WPDD}\left ( S;k \right )\in \mathbb {R} ^{n\times \left ( k+1 \right )}$ is formed, represented by the following equation:
\begin{equation}
\begin{array}{@{}c@{}c@{}}
    \resizebox{.91\linewidth}{!}{$\mathcal{WPDD}=\left ( \mathrm{W}, \mathcal{PDD} \right )=  \left({ \bigcup_{i=1}^{n}} \frac{t(x_{i})}{ \sum_{u=1}^{n} t(x_{u}) },  {\bigcup_{j=1}^{k}} d(p_{i},p_{j}) \right)$}
\end{array}
\end{equation}
Herein, $n$ represents the number of atoms in the unit cell, and $p_{i}$ and $p_{j}$ denote the spatial coordinates of an atom $i$ and its neighbor $j$, respectively,  and $k$ is the number of nearest neighbors selected when constructing the PDD, sorted in ascending order of Euclidean distance as $d_{i1} \le \dots \le d_{ik}$, as shown in Figure \ref{3DPDD}(b). WPDD is equivalent to the PDD of the crystal structure $\mathcal{S}$, except that the rows are not grouped as in the original version. This prevents the loss of atomic information when two primitive points have the same $k$-nearest neighbor distances but correspond to different atomic types. Therefore, $\mathcal{WPDD} \in\mathbb{R}^{n \times \left (k+1 \right ) }$, where $n$ is the number of atoms in the constructed graph.

\subsection{Unit-cell PDD (UPDD)}
When ensuring the completeness of PDD, a large number of neighbors must be predefined, typically requiring information on hundreds of neighbors, and in extreme cases, the number must exceed the atom count in any unit cell within the dataset. The number of neighbors, $k$, is difficult to determine across different datasets, and for unit cells with fewer atoms, which constitute a larger proportion of the dataset, an excess of neighbor information may interfere with the speed of message aggregation, leading to greater resource consumption.

To address this issue, we introduce Unit-cell PDD (UPDD). We achieve this by reconstructing the unit cell around each atom and encoding the pairwise distances between the atom and other atoms within the reconstructed unit cell. 
This means that when constructing PDD, we focus more on the overall structure of the atoms within the reconstructed unit cell, thereby reducing interference from excessive neighbor information. 
UPDD is defined by the following formula:
\begin{equation}
\begin{array}{@{}c@{}c@{}}
\resizebox{.91\linewidth}{!}{$\mathcal{UPDD} = \left \{ \bigcup_{i=1}^{n}  {\bigcup_{j=1}^{n}} d(p_{i},p_{j})^{-1}  \mid i, j\in \mathcal{Z} ,p_{i} \ne p_{j} \right \}$}
\end{array}
\end{equation}
Since the interaction energy between an atom and its neighboring atoms is usually inversely proportional to the distance, we take the reciprocal feature of the distance after removing zeros.

As shown in Figure \ref{3DPDD}(d), the selection is not based on Euclidean distances but on choosing atoms within the reconstructed unit cell for construction. 
The atoms in the unit cell determine the dimension of our UPDD and do not require consideration of the neighbor count, k, across different datasets, making it more generalizable. This UPDD covers unit cell structures with a larger number of atoms while ensuring that unit cell structures with fewer atoms are not disturbed by excessive neighbor information. It also reduces resource consumption. Due to this crystal-specific treatment, the UPDD dimensions of different crystal structures may not match, so dimension alignment is required before feeding them into the neural network.

\subsection{Crystal Graph Construction}
\label{Crystal_graph}
By introducing PDD, we constructed a general complete and continuous multi-edge crystal graph. 
In the graph, node features are $x_{i}$. 
An edge is established from node $j$ to node $i$ when the Euclidean distance $\left | e_{j'i} \right | ^{2}$ between a duplicate of $j$ and $i$ satisfies $\left | e_{j'i} \right | ^{2} = \left | p_{j} + k'_{1} l_{1} + k'_{2} l_{2} + k'_{3} l_{3} - p_{i} \right | ^{2} \le r$, 
where $r \in \mathbb{R}$ is the cutoff radius. 
Next, we construct a PDD row for each atom. Since directly representing PDD as edge features is impractical, we retain its matrix form and incorporate it into the construction of the multi-edge crystal graph to reflect the global information of the crystal structure.
Therefore, we represent the constructed crystal graph as $\mathcal{G} = (\mathcal{X}, \mathcal{XI}, \mathcal{E}, \mathcal{PDD})$. 
Therein, $x_{i} \in \mathcal{X}$ is the feature vector of the atom $i$, $e_{ij}^{h} \in \mathcal{E}$ is the feature vector of the $h$-th edge between nodes $i$ and $j$, and we denote $\mathcal{XI}$ as the indices of the nodes $i$ and $j$ that form the edge. Sections \ref{Continuity} and \ref{completeness} are our proofs of the continuity and general geometric completeness of PDD crystal graphs.
\begin{figure*}[t]
\centering
\includegraphics[width=0.92\textwidth]{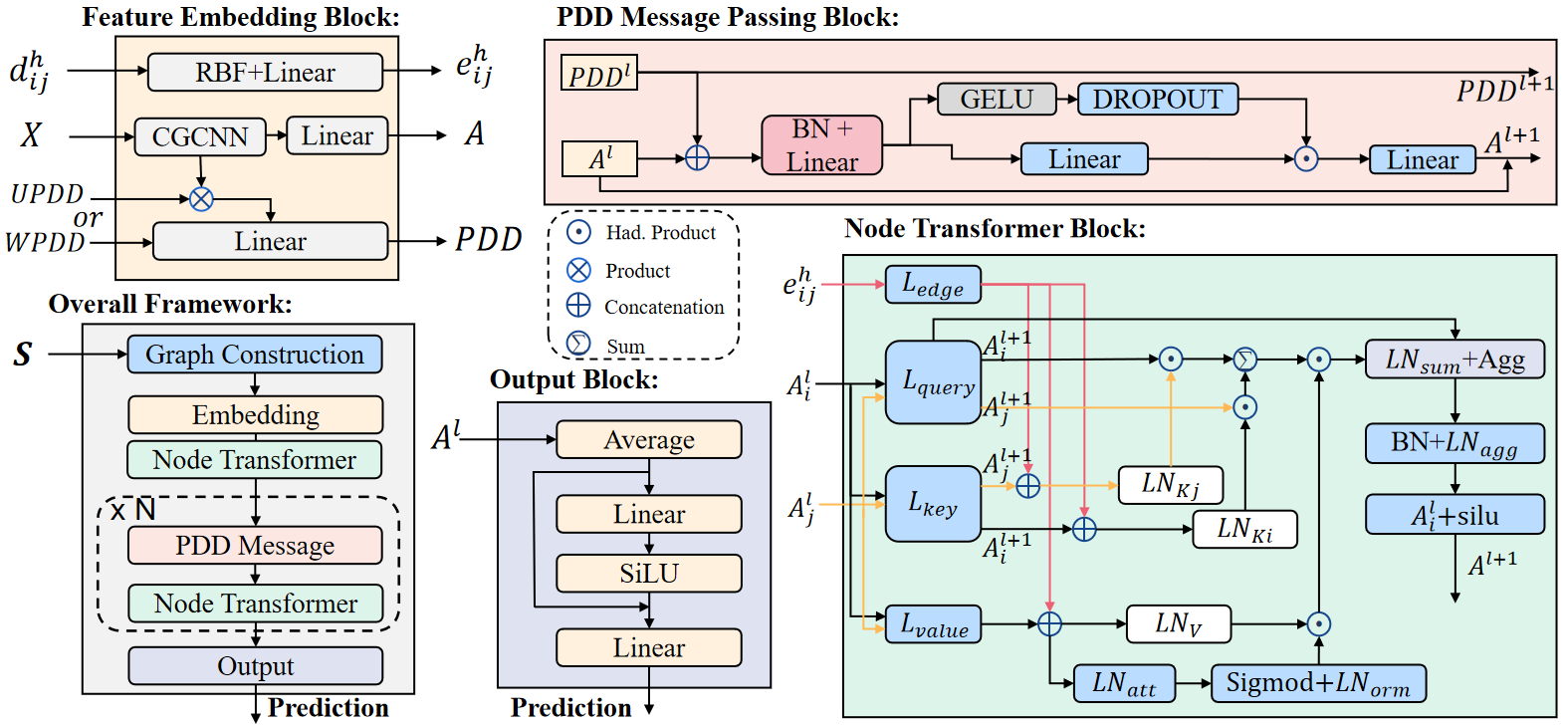} 
\caption{Architecture Overview. PDDFormer accepts an input crystal structure $S$. During the prediction process, it first undergoes a graph construction step to generate a continuous and general complete crystal graph structure, followed by an embedding block, then multiple blocks of node-wise Transformer and PDD Message Passing, and finally, an output block.}
\label{fig3}
\end{figure*}

\subsection{The Continuity of Proposed Crystal Graphs}
\label{Continuity}
The continuity of the constructed crystal graph $\mathcal{G}\left ( \mathcal{S}  \right )$ under perturbations of the crystal structure $\mathcal{S}$ will be measured using the EMD~\cite{rubner2000earth}, which applies to crystal graphs of any size. 
Definition 6 applies to any crystal graph $\mathcal{G} \left ( \mathcal{S}  \right ) =\left ( \left [  w_{1}\left ( S \right ), R_{1}\left ( S \right )  \right ] , \dots ,\left [  w_{n}\left ( S \right ), R_{n}\left ( S \right )  \right ] \right )$, where $\left [  w_{i}\left ( S \right ), R_{i}\left ( S \right )  \right ]$ represent the information extracted based on atom $i$ in the unit cell. $R_{i}\left ( S \right )= R_{i}\left ( \mathcal{S}_{\mathcal{X}}, \mathcal{S}_{\mathcal{XI}}, \mathcal{S}_{\mathcal{E}}, \mathcal{S}_{\mathcal{PDD}} \right )$ includes atomic information, neighbor information used in constructing the multi-edge crystal graph, and the PDD invariants of the crystal structure $\mathcal{S}$, with weights $w_{i} \in \left ( 0,1 \right ] $ satisfying the normalization condition $ {\textstyle \sum_{i=1}^{n}} w_{i} \left ( \mathcal{S}  \right ) =1$. 

Subsequently, we only consider the case where the weighted vector $\left [ w_{i}, R_{i}  \right ] $ corresponds to the $i$-th row of $\mathcal{PDD}\left ( S;k \right ) $. Here, $n$ denotes the number of rows in $\mathcal{PDD}\left ( S;k \right ) $. The size of each row $R_{i} \left ( S \right ) $ should be independent of $\mathcal{S}$ and depend solely on the number of neighbors $k$ in $\mathcal{PDD}\left ( S;k \right ) $. For any vectors $R_{i} = \left ( r_{i1} ,\dots ,  r_{ik}\right ) $ and $R_{j} = \left ( r_{j1} ,\dots ,  r_{jk}\right ) $ of length $k$, we use the $L_{\infty } $- distance $\left | R_{i} -R_{j}  \right | _{\infty } = \max_{l = 0,\dots k} \left | r_{il} -r_{jl}  \right | _{\infty }$.

\begin{proposition}
The WPDD and UPDD multi-edge crystal graph is continuous.
\end{proposition}
\begin{proof}
For any neighbor count $k \geq 1$, if the  periodic crystal $\mathcal{S}, \mathcal{Q}\in \mathbb{R}^{n}$ satisfy $d_{B}\left ( \mathcal{S}, \mathcal{Q}  \right )<r\left ( \mathcal{S} \right )$, then we have:
\begin{equation}
\begin{aligned}
\mathrm{EMD}\left ( \mathcal{G}  \left ( \mathcal{S}  \right ), \mathcal{G}\left ( \mathcal{Q}  \right ) \right ) &=\\
\mathrm{EMD}\left ( \left ( \mathcal{S}_{\mathcal{X}}, \mathcal{S}_{\mathcal{XI}}, \mathcal{S}_{\mathcal{E}}, \mathcal{S}_{\mathcal{PDD}} \right ), 
\left ( \mathcal{Q}_{\mathcal{X}}, \mathcal{Q}_{\mathcal{XI}}, \mathcal{Q}_{\mathcal{E}}, \mathcal{Q}_{\mathcal{PDD}} \right )  \right ) 
&= \\
\mathrm{EMD}\left ( \left ( \mathcal{S}_{\mathcal{X}}, \mathcal{Q}_{\mathcal{X}} \right )  \right ) +
\mathrm{EMD}\left ( \left ( \mathcal{S}_{\mathcal{XI}}, \mathcal{Q}_{\mathcal{XI}} \right )  \right ) 
&+\\ \mathrm{EMD}\left ( \left ( \mathcal{S}_{\mathcal{E}}, \mathcal{Q}_{\mathcal{E}} \right )  \right )+
\mathrm{EMD}\left ( \left ( \mathcal{S}_{\mathcal{PDD}}, \mathcal{Q}_{\mathcal{PDD}} \right ) \right ).
\end{aligned}
\end{equation}
Since disturbances only change the positions of atoms and do not alter their types, therefore $\mathrm{EMD}\left ( \left ( \mathcal{S}_{\mathcal{X}}, \mathcal{Q}_{\mathcal{X}} \right )  \right ) = 0$ and $\mathrm{EMD}\left ( \left ( \mathcal{S}_{\mathcal{XI}}, \mathcal{Q}_{\mathcal{XI}} \right )  \right )=0$, refer to $\mathit{Appendix}$ F for details. So, we obtain $\mathrm{EMD}\left ( \mathcal{G}\left ( \mathcal{S}  \right ), \mathcal{G}\left ( \mathcal{Q}  \right ) \right )= \mathrm{EMD}\left ( \left ( \mathcal{S}_{\mathcal{E}}, \mathcal{Q}_{\mathcal{E}} \right )  \right )+\mathrm{EMD}\left ( \left ( \mathcal{S}_{\mathcal{PDD}}, \mathcal{Q}_{\mathcal{PDD}} \right )  \right ) \le 2d_{B}\left ( \mathcal{S},  \mathcal{Q}  \right )$.
\end{proof}

The bottleneck distance $d_{B}\left ( \mathcal{S},  \mathcal{Q}  \right )<r\left ( \mathcal{S} \right )$ is defined as: $d_{B}\left ( \mathcal{S}, \mathcal{Q}\right )= \inf_{g:\mathcal{S } \to \mathcal{Q} } \sup_{p \in \mathcal{S }}\left | p-g\left ( p \right )  \right | $ and the envelope radius $r\left ( \mathcal{S} \right )$ is the minimum half-distance between any two points in $r\left ( \mathcal{S} \right )$. In other words, $r\left ( \mathcal{S} \right )$ is the maximum radius of non-overlapping open balls centered at all points in $\mathcal{S}$. This implies that any small perturbation in atomic positions under the $d_{B}$~\cite{carstens1999geometrical} will lead to minor changes in the distribution of distances between points in the EMD.
 
Since the EMD between the constructed crystal graphs only relates to Euclidean distance. Euclidean distance itself is continuous, Theorem 1 extends the following fact: for a unit cell structure with two atoms, when the number of neighbors $k = 1$, if we perturb at most two points by $\epsilon$, the change in distance between the two points will be at most $2\epsilon$. Extending this to $n$ atomic points with $k$ neighbors, if we perturb at most $n$ points by $\epsilon$, the change in distance between $n$ points will be at most $2nk\epsilon$. 
This aligns with Definition 5, hence, the constructed WPDD and UPDD multi-edge crystal graph is continuous.

\begin{table*}[htbp]
  \centering
  \resizebox{\textwidth}{!}{  
  \begin{tabular}{lccccccccc}
    \toprule
    \multirow{2}{*}{\textbf{Method}} & \multicolumn{1}{c}{Formation Energy} & \multicolumn{1}{c}{Bandgap(OPT)} & \multicolumn{1}{c}{Total Energy} & \multicolumn{1}{c}{Ehull} & \multicolumn{1}{c}{Bandgap(MBJ)}  &\multicolumn{1}{c}{Bulk Moduli(Kv)} & \multicolumn{1}{c}{Shear Moduli(Gv)} &
    \\
    \cmidrule(r){2-8}
     & eV/atom &  eV &  eV/atom & eV & eV  &GPa &  GPa
    \\
    \midrule
    PotNet(2023) & 0.0294  & 0.127 & 0.032 &  0.055 &0.27 & 10.06 & 8.883 \\
    CrysMMNet(2023) & 0.028  & 0.128 & 0.034 & --  & 0.278 & 9.625  & \underline{8.471}  \\
    CrysDiff (2024) & 0.029  & 0.131 & 0.034 & 0.062 &  0.287 & 9.875  & 9.193 \\
    Crystalformer(2024)  & 0.0306  & 0.128 & 0.032 & 0.046 &  0.274 & -- & -- \\
    eComFormer(2024)  & 0.0284  & 0.124 & 0.032 & 0.044 &  0.28 & 10.79  & 9.826\\
    iComFormer(2024) & {0.0272}  & \underline{0.122} & {0.0288} & 0.047 &  {0.26}& {9.617}  & 9.098 \\
    
    \midrule
        \textbf{UPDDFormer}   & \underline{0.0267} & \underline{0.122} & \underline{0.0287} & \underline{0.0406} &  \underline{0.260} & \underline{9.456} & {8.738}  \\
    \textbf{WPDDFormer}   & \textbf{0.0257} & \textbf{0.119} & \textbf{0.0276} & \textbf{0.0355} &  \textbf{0.249}  & \textbf{9.224} & \textbf{8.441} & \\
    \bottomrule
  \end{tabular}
  }
    \caption{Comparison between UPDDFormer, WPDDFormer, and other baselines in terms of test MAE on the JARVIS dataset. The best results are shown in \textbf{bold} and the second-best results are shown with \underline{underlines}.}
    \label{sample-table1}
\end{table*}

\begin{table}[htbp]
    \centering
    \renewcommand{\arraystretch}{1.2}
    \resizebox{\linewidth}{!}{
    \begin{tabular}{lccccc}
        \toprule
        \multirow{2}{*}{\textbf{Method}} & \multicolumn{1}{c}{Formation Energy} & \multicolumn{1}{c}{Band Gap} & \multicolumn{1}{c}{Bulk} & \multicolumn{1}{c}{Shear} \\
        \cmidrule(r){2-5}
            & eV/atom &  eV &   log(GPa) &  log(GPa)   \\
        \midrule
        PotNet   & 0.0188  & 0.204 & 0.040 &  0.065   \\
        CrysMMNet   & 0.0200  & 0.197 & 0.038 &  \underline{0.062}   \\
        Crystalformer  & 0.0186 & 0.198 & {0.0377} & 0.0689 \\
        eComFormer   & {0.01816}  & 0.202 & 0.0417 & 0.0729  \\
        iComFormer & 0.01826  & \underline{0.193} & 0.0380 & 0.0637  \\
        \midrule
        \textbf{UPDDFormer}   & \underline{0.01696} & \underline{0.189} & \underline{0.0370} & {0.0670}  \\
        \textbf{WPDDFormer}   & \textbf{0.01604} & \textbf{0.187} & \textbf{0.0323} & \textbf{0.0605}  \\
        \bottomrule
    \end{tabular}
    }
        \caption{Comparison of test MAE between UPDDFormer, WPDDFormer, and other baselines on the Materials Project dataset.}
    \label{sample-table2}
\end{table}

\subsection{General Geometric Completeness}
\label{completeness}
\begin{proposition}
The WPDD multi-edge crystal graph is generally geometrically complete.
\end{proposition}

We discuss the limitations of general completeness in $\mathit{Appendix}$ G. We prove this by categorical induction. Crystal structures can be classified into stable and unstable types and further divided into the following three categories:
1. Stable Crystal Structures(i.e., where no two crystals can have the same structure with only a difference in atomic types).
2. Unstable Crystal Structures with Differences in Atomic Coordinates.
3. Unstable Crystal Structures with Identical Structures but Differences in Atomic Types. 

\begin{proof}
Since UPDD is constructed based on the size of the unit cell, when the number of atoms in the unit cell is relatively small, it could theoretically result in different crystal structures, where all atoms have the same Euclidean distances and atom types but inconsistent atomic positions, sharing the same crystal graph representation. 

Since the crystal structures of the first and second categories differ, PDD alone can effectively distinguish them. As WPDD includes PDD, different WPDD representations can be constructed for the two crystals, thereby achieving differentiation.
For the third category, we incorporate WPDD as global information into the construction of a multi-edge crystal graph and encode atomic information, such that the WPDD crystal graph $\mathcal{G}$ is represented as $\mathcal{G} = (\mathcal{X}, \mathcal{XI}, \mathcal{E}) + \text{WPDD}$, where $\mathcal{X}$ represents atomic information embedded through CGCNN. 
For any atom in the unit cell, a WPDD row vector needs to be constructed along with the corresponding atomic information embedding. This ensures that for any two crystal structures with identical crystal structures but differing atomic types at corresponding coordinates, the $(\mathcal{X}, \mathcal{XI}, \mathcal{E})$ in their WPDD crystal graphs will differ. 
On the contrary, if two crystals have the same WPDD crystal graph representation, they share the same WPDD and multigraph representations. It indicates that their crystal structures and the atomic information at corresponding coordinates are identical, thus confirming that they are the same crystal. This contradicts our premise. Hence, the proof is complete. 
Therefore, the proposed identical crystal graph can represent only the same infinite crystal structure. Then, based on Definition 3, we complete the proof of Proposition 2.

Finally, we conclude that the UPDD crystal graph can only guarantee continuity, while the WPDD crystal graph can ensure both continuity and general completeness.
\end{proof}

\subsection{Network Architecture}
\label{network}
Based on the graph in Section \ref{Crystal_graph}, we propose the information propagation scheme of PDDFormer. Figure \ref{fig3} illustrates the overall framework architecture of PDDFormer.

\paragraph{Feature Embedding Block} First, we introduce the construction of the graph embedding Block. We use atomic encoding from CGCNN for embedding. For the edge information $e_{ij}^{h}$, we employ radial basis functions to encode the distance between two adjacent nodes in the graph, represented by Equation \ref{eq2}, where $\gamma$ and $\mu$ are hyperparameters. 
For UPDD, due to the varying feature dimensions of UPDD for different crystals, we perform matrix multiplication on UPDD to align the structural information of different crystals, obtaining information for the PDD message passing layer. Thus, we obtain the graph embedding as:
\begin{equation}
\begin{array}{@{}c@{}c@{}c@{}}
\label{eq2}
    \mathrm{A} = {\mathrm{CGCNN}}(\mathcal{X} ), \quad
    e_{ij}^{h} = \exp\left(-\gamma \left(\frac{\left\| p_{i}-p_{j} \right\|^2}{\mu}\right)\right), \: \\
    \mathcal{PDD} = \mathcal{UPDD} \otimes \mathrm{A} \quad or \quad  \mathcal{WPDD}
\end{array}
\end{equation}

\begin{table*}[th]
  \centering
  \begin{tabular}{ccccccc}
   \toprule
    Models     & Time/epoch &  Total time & Inference time & GPU memory usage &Complexity & Model Parameter   \\
    \midrule
      eComformer & 97s  &13.4h & 361.5s &17GB& $O(nk)$&12.4M\\
      iComformer & 103s  &14.3h &365.6s &12GB& $O(nk)$&5.0M\\
       \midrule
    \textbf{WPDDFormer}    & 66s &7.4h &240.6s  &6GB& $O(nk)$&4.52M\\
    \bottomrule
  \end{tabular}
        \caption{Efficiency comparison with  ComFormer on the Jarvis Ehull task. We show the training time per epoch, inference time for the whole test set, total training time, time complexity, GPU memory consumption, and total number of parameters. The experiments were conducted using a 3090 RTX 24GB GPU. The data in the table is averaged over three experiments.}
\label{efficency}
\end{table*}
\begin{table}[t]
\centering
\begin{tabular}{cccc}
    \toprule
        Method &Num. Block & Ehull  &Band Gap\\
    \midrule
    NO PDD Block  &3,0&0.0430 &0.194  \\
    NO PDD &3,2  & 0.0426 &0.193 \\
    \midrule
    \textbf{UPDDFormer}  &3,2 & {0.0406} &  {0.189} \\
    \textbf{WPDDFormer} &3,2 & \textbf{0.0355} &  \textbf{0.187} \\
    \hline
    \label{ablation}
\end{tabular}
\caption{Num. Block represents the number of Node transformer blocks and PDD message passing blocks.}
\label{table2}
\end{table}
\paragraph{Node Transformer Block} 
Building upon the constructed graph, we aggregate the node information. 
Let $a_{i}^{l} $ be the input feature vector of node $i$ at layer $l$ in PDDFormer. 
The information propagation of layer $l$ is formulated as follows:
\begin{equation}
\begin{array}{@{}c@{}c@{}c@{}}
k_{i}^{l} = \left ( LK\left ( a_{i}^{l}   \right )\oplus LE\left ( e_{ij}^{h}   \right )   \right ) ,
    k_{j}^{l} = \left ( LK\left ( a_{j}^{l}   \right ) \oplus LE\left ( e_{ij}^{h}   \right )   \right ), \: \\
    v = \left ( LV\left ( a_{i}^{l}   \right ) \oplus LV\left ( a_{j}^{l}   \right ) \oplus LE\left ( e_{ij}^{h}   \right )   \right ), \:q_{i}^{l} = LQ\left ( a_{i}^{l}   \right ) ,  \\
    q_{j}^{l} = LQ\left ( a_{j}^{l}   \right ) ,v_{ij}^{l} = v \odot \mathrm{Sigmoid}\left ( LN_{orm}\left (LN_{att}\left ( v \right ) \right) \right ),  \: \\
    att^{l} =  \frac{q_{i}^{l} \odot  LN_{Kj}\left ( k_{j}^{l} \right ) + q_{j}^{l} \odot  LN_{Ki}\left ( k_{i}^{l} \right ) }{\sqrt{d_{q_{i}^{l}} }}, \: \\
    m_{ij}^{h} = q_{i}^{l}+\mathrm{Sigmoid}\left ( \mathrm{BN}\left  ( att^{l} \right) \right ) \odot LN_{V}\left ( v_{ij}^{l} \right )
\end{array}
\end{equation}
where $LQ, LK, LV, LE$ are the linear transformations for query, key, value, and edge features. $LN_{K}$, $LN_{V}$ are the nonlinear transformations for key and value, including two linear layers and an activation layer in between.  $LN_{att}$ represents the linear transformation for updating messages, and $LNorm$ denotes the layer normalization~\cite{ba2016layer} operation. BN denotes the batch normalization layer~\cite{ioffe2015batch}, and $d_{q_{i}^{l}}$ is the dimension of $q_{i}^{l}$. 

Then, we obtain the message $M_{i}^{l}$ by aggregating the information from the neighborhood of node $i$ over multiple edges, and $A_{i}^{l+1}$ is realized as follows :
\begin{equation}
\begin{array}{@{}c@{}c@{}}
    M_{i}^{l} =  \mathrm{BN}\left (\sum_{j\in A_{i} } \sum_{h} LN_{sum}\left ( m_{ij}^{h} \right ) \right ) , \: \\
    A_{i}^{l+1} = \mathrm{SiLU}\left ( a_{i}^{l}  +  LN_{agg}\left (M_{i}^{l} \right )   \right )
\end{array}
\end{equation}
where $LN_{sum}$ is the linear transformation used for updating the edge messages.

\paragraph{PDD Message Passing Block} 
$\mathcal{A} ^{l}$ and $PDD^{l}$ represent the atomic features and 3D periodic pattern encoding at layer $l$, respectively. Its message-passing mechanism is as follows:
\begin{equation}
\begin{array}{@{}c@{}c@{}c@{}}
    \resizebox{.9\linewidth}{!}{$     \mathcal{PDD}^{l+1} = \mathcal{PDD}^{l}+\mathcal{A} ^{l+1} , \:
    A_{1}, A_{2} = LN_{\mathcal{PDD}}\left ( \mathrm{BN}\left ( \mathcal{PDD}^{l+1} \right )  \right ) , $}\\
    \resizebox{.9\linewidth}{!}{$  A^{l+1} = A^{l}+  LN_{A2}
    \left (LN_{A1}\left ( A_{1} \right ) \odot  \mathrm{Drop}\left (\mathrm{GELU}\left ( A_{2} \right ) \right ) \right )  $}
\end{array}
\end{equation}
Finally, we use average pooling to aggregate the features of all nodes in the graph, followed by a nonlinear layer, and then a linear layer to obtain the scalar output of the graph as described above. A detailed description of the PDDFormer architecture can be found in $\mathit{Appendix}$ B.

\section{Experiments}
We conducted experiments on two material benchmark datasets, namely the Materials Project~\cite{chen2019graph} and Jarvis~\cite{choudhary2020joint} datasets. 
Detailed descriptions of the datasets can be found in $\mathit{Appendix}$ A. More information about the experimental settings of PDDFormer can be found in $\mathit{Appendix}$ C. Baseline methods include PotNet~\cite{lin2023efficient}, CrysMMNet~\cite{das2023crysmmnet}, CrysDiff~\cite{song2024diffusion}, Crystalformer~\cite{taniai2024crystalformer}, and ComFormer~\cite{yan2024complete}, The complete baselines can be found in $\mathit{Appendix}$ E.1.
For all baselines on the datasets, we report the results provided in the cited papers.

\subsection{Experimental Results}

\paragraph{JARVIS} 
The quantitative results for JARVIS~\cite{choudhary2020joint} are shown in Table~\ref{sample-table1}. WPDDformer achieves the best performance across all tasks. UPDD achieved the second-best results. Notably, WPDDFormer outperforms eComFormer by 19\% respectively in the Ehull task. 

\paragraph{The Materials Project (MP)} 
The experimental results on MP~\cite{chen2019graph} are shown in Table~\ref{sample-table2}. WPDDFormer significantly outperforms previous works across all tasks, with a $11.7$\% improvement over the second-best model in the critical formation energy task and a $14.3$\% improvement in the bulk moduli task. Additionally, the excellent prediction accuracy of WPDDFormer in the bulk modulus and shear modulus tasks, using only $4,664$ training samples, demonstrates the expressiveness and robustness of WPDD multi-edge crystal graphs under limited training samples.
Overall, our methods are compared with $14$ existing methods across the two datasets. Our WPDDFormer consistently outperforms all methods in all tasks. Additionally, WPDDFormer shows a significant improvement in prediction accuracy compared to UPPDFormer. This improvement is not only because the WPDD graph structure is complete and continuous, while UPPD can only ensure continuity, but also because UPPD requires dimensional alignment as mentioned in \ref{network}, which results in some loss of the expression of global information about the unit cell. 
\paragraph{Efficiency}
We compare model efficiency with Comformer under the best configurations, as reported in Table~\ref{efficency}. 
They have a time complexity of $O(nk)$, where $n$ is the number of atoms in the unit cell and $k$ is the average number of neighbors.
Compared to ComFormer, WPDDFormer has fewer parameters and achieves significantly superior experimental results with lower computational cost and faster training and inference speeds, demonstrating the remarkable superiority of our method.

\subsection{Ablation Studies}
In this section, we demonstrate the impact of introducing (W/U)PDD on the representation learning of crystal materials through ablation studies. Specifically, we conducted experiments on the MP and JARVIS datasets, using testing mean absolute error (MAE) as the quantitative evaluation metric, comparing the results for \textbf{Band Gap} and \textbf{Ehull} tasks, as shown in Table~\ref{table2}. 
By comparing (W/U)PDDFormer models without PDD message passing blocks to models that retain the ${PDD}$ message passing blocks but lack (W/U)PDD information, we validated the importance of (W/U)PDD. The results show that compared to models without the $PDD$ message passing blocks, WPDDForemer achieved improvements of $17.4\%$ and $3.6\%$ in the Bulk Moduli and Ehull tasks, respectively. Compared to models that retain only the $PDD$ message passing blocks but lack (W/U)PDD information, we achieved improvements of $16.7\%$ and $3.1\%$ in these two tasks, respectively.  More data can be found in $\mathit{Appendix}$ E.3.

\section{Conclusion}
In this study, we integrated WPDD and UPDD into the representation of crystal structures, achieving a general complete and continuous construction of crystal graphs. 
This resolves the ambiguity in crystal graph representations for predicting the properties of crystalline materials and bridges the gap between traditional crystal descriptors and dynamic atomic behavior. 
Experimental results demonstrate the significant advantage of our PDDFormer in various property prediction tasks.
Achieving absolute completeness under perturbations is a problem that will be further explored in the future.

\section*{Acknowledgments}
This research is supported by the National Natural Science Foundation of China (No.42130112, No.42371479), General Program of Shanghai Natural Science Foundation(Grant No.24ZR1419800, No.23ZR1419300), Science and the Technology Commission of Shanghai Municipality (Grant No.22DZ2229004), Beijing Natural Science Foundation (No.QY23187), and Shanghai Frontiers Science Center of Molecule Intelligent Syntheses.

\section*{Contribution Statement}
M.C., X.T., X.H., J.Y., and X.W. supervised the research project throughout its duration. Z.W. and X.S. conceived the experimental design and jointly developed the model framework. M.C., X.T., X.H., J.Y., S.L., and X.W. provided revisions to the manuscript. The experimental results were collaboratively analyzed by L.S., L.W., Z.W., and X.S. Data interpretation and manuscript writing were carried out by Z.W. and X.S. X.S. and Z.W. are co-first authors, and X.W. is the corresponding author. All authors participated in the discussion of the results and contributed to the final manuscript preparation.

\bibliographystyle{named}
\bibliography{ijcai25}

\appendix
\newpage
\section{Dataset descriptions}
\label{dataset}
In this section, we provide more detailed information about the JARVIS and The Materials Project datasets.
\paragraph{The Materials Project dataset.}
Materials Project (MP) is a collection of $69,239$ materials from the Materials Project database
retrieved by~\cite{chen2019graph}.  We follow the experimental setup of Matformer ~\cite{yan2022periodic} using the same training, validation, and test sets. For the formation energy and band gap tasks, the training, validation, and test sets contain crystals of $60,000$, $5,000$, and $4,239$, respectively. 
Among these, there are 38,344 samples with at least $20$ atoms per unit cell, accounting for approximately $55.4$\%. There are $2,047$ samples with at least 100 atoms per unit cell, accounting for about $3.0$\%. We evaluate our Matformer on four key crystal property tasks: formation energy, band gap, bulk modulus, and shear modulus. For the bulk modulus and shear modulus tasks, the training, validation, and test sets contain $4,664$, $393$, and $393$ crystals, respectively. 
\paragraph{The JARVIS dataset.} 
JARVIS is a database proposed by Choudhary et al. ~\cite{choudhary2020joint}. For the JARVIS dataset, we follow the approach of Matformer ~\cite{yan2022periodic} and divide the data into training, validation, and test sets in an $8:1:1$ ratio. We evaluate our PDDFormer model on nine crucial crystal property tasks, including formation energy, bandgap (OPT), bandgap (MBJ), total energy, Ehull, bulk modulus (Kv), shear modulus (Gv), SLME (\%), and Spillage. For the formation energy, total energy, and bandgap (OPT) tasks, the training, validation, and test sets contain $44,578$, $5,572$, and $5,572$ crystal samples, respectively. Among these, there are 8,089 samples with at least $20$ atoms per unit cell, accounting for approximately $14.5$\%. Only $4$ samples have at least $100$ atoms per unit cell. For the Ehull task, these numbers are $44,296$, $5,537$, and $5,537$ samples; for the bandgap (MBJ) task, they are $14,537$, $1,817$, and $1,817$ samples; for bulk modulus (Kv) and shear modulus (Gv) tasks, they are $15,744$, $1,968$, and $1,968$ samples; for SLME (\%) task, they are $7,254$, $906$, and $906$ samples; and for the Spillage task, they are $9,101$, $1,137$, and $1,137$ samples.


\begin{table*}[th]
  \caption{Model settings of WPDDFormer for JARVIS dataset.}
  \centering
  \begin{tabular}{ccccc}
   \toprule
    Parameter     & Learning rate &  Num. neighbors &  Epoch number &Num. Node and PDD   \\
    \midrule
    formation energy &  0.001 &  25 &  400  &3,2   \\
    band gap (OPT) & 0.0005  & 25 & 400 &    3,2  \\
    total energy & 0.001  & 25 & 400 &3,2    \\
    Ehull & 0.001  & 25 & 400 &3,2    \\
    band gap (MBJ) & 0.001  & 25 & 300 &    3,2  \\
    Bulk Moduli(Kv) & 0.001  & 18 & 300 &3,2     \\
    Shear Moduli(Gv)  & 0.001  & 18 & 300&3,2    \\
    SLME(\%) & 0.001  & 18 & 300 & 3,2     \\
    Spillage & 0.001  & 12 & 200&3,2     \\
    \bottomrule
  \end{tabular}
\label{JARVIS_Para}
\end{table*}

\begin{table*}[th]
  \caption{Model settings of WPDDFormer for The Materials Project dataset.}
  \centering
  \begin{tabular}{ccccc}
   \toprule
    Parameter     & Learning rate &  Num. neighbors &  Epoch number &Num. Node and PDD   \\
    \midrule
    formation energy &  0.001 &  25 &  500  &3,2   \\
    band gap & 0.001  & 25 & 400 &    3,2  \\
    bulk moduli & 0.001 & 16 & 300   &3,2   \\
    shear moduli & 0.001  & 16 & 300 &3,2    \\
    \bottomrule
  \end{tabular}
\label{MP_Para}
\end{table*}

\section{PDDFormer configurations}
\label{detail pddformer}
We trained on the JARVIS and the MP datasets using an RTX 4090 24GB GPU.
\paragraph{Notations.}
$\mathcal{A} \in \mathcal{R} ^{n\times da}$ is the atomic feature matrix obtained by embedding the atomic matrix $\mathcal{X} \in \mathcal{R} ^{n\times 1} $ in the unit cell, where n represents the number of atoms in the unit cell, $\mathcal{A} = \left [ a_{1}, a_{2}  \cdots a_{n-1}, a_{n}\right ]^{T}\in \mathrm {R} ^{n\times da}$, and $a_{i}$ represents the da-dimensional feature vector of atom $i$ in $\mathcal{A}$. $e_{ij}^{h} \in \mathcal{E}$ is the de-dimensional feature vector of the h-th edge connecting nodes $i$ and $j$. Typically, de is the same dimension as da.
In constructing the PDD, we take the nearest neighbors $k = 92$, resulting in its dimensional information where PDD $\in \mathcal{R}^{n \times 92}$. The WPDD incorporates an additional dimension for atomic weights, thus its dimension is WPDD$ \in \mathcal{R}^{n \times 93}$. For the UPDD, prior to alignment, its dimension is solely related to the number of atoms in the unit cell, expressed as UPDD$\in \mathcal{R}^{n \times n}$. After alignment, it matches the dimension of $\mathcal{A}$ to facilitate information aggregation. $\mathcal{XI} \in \mathcal{R} ^{tn\times 2}$ is the index of the points corresponding to the edge, where $t$ is the number of the nearest edges aggregated within our cutoff radius  ~\cite{flor2016comparison}.
\paragraph{Graph embeddings.}
For the two datasets we use, we employ the CGCNN atomic embedding, where the atomic number is mapped to a $92$-dimensional embedding vector. Subsequently, we apply a linear transformation to map it to a $256$-dimensional vector, serving as the input $a_{i}$ passed to the first PDDFormer message update block. For each edge, we employ $256$ RBF kernels to map the Euclidean distance to a $256$-dimensional embedding vector, with kernel centers ranging from $-6.0$ to $0.0$. It is then mapped to a $256$-dimensional vector as the edge input $e_{ij}$, through a nonlinear layer followed by a linear layer. For the UPDD, after performing matrix multiplication with the embedded atomic information $\mathcal{A}$, it is passed through a nonlinear layer and a linear layer to map it to the same dimension as $\mathcal{A}$. For the PDD and WPDD, they are directly processed through a nonlinear layer and a linear layer to map them to the same dimension as $\mathcal{A}$ and then passed to the message update layer.

\paragraph{Settings of node transformer block.} LQ, LK, LV, and LE are linear transformation layers that map $256$-dimensional input features to $256$-dimensional output features. $LN_{att}$ and $LN_{orm}$ are linear transformation layers that map $256 \times 3$ dimensional input features to $256$ dimensional output features and layer normalization ~\cite{ba2016layer}, respectively. $LN_{V}$ is nonlinear transformations for value, including one linear layer that maps the concatenated $256\times3$ dimensional input features to $256$-dimensional output features, one SiLU activation layer ~\cite{paul2022sinlu}, and one linear layer that maps the $256$-dimensional input features to $256$-dimensional output features. $LN_{ki}$ and $LN_{kj}$are nonlinear transformations for key, including one linear layer that maps the concatenated $256\times2$ dimensional input features to $256$-dimensional output features, one SiLU activation layer ~\cite{paul2022sinlu}, and one linear layer that maps the $256$-dimensional input features to $256$-dimensional output features.

\paragraph{Settings of PDD message passing block.} $LN_{PDD}$ is a linear transformation layer that maps $256$ dimensional input features to $256$-dimensional output features. Then, the first $128$ dimensions are assigned to A1, and the remaining dimensions are assigned to A2. Dropout ~\cite{srivastava2014dropout} is set to $0.1$. After passing through $LN_{A1}$, which is a linear transformation layer that maps $128$ dimensional input features to $128$ dimensional output features, the data then goes through $LN_{A2}$, another linear transformation layer that maps $128$ dimensional input features to $256$ dimensional output features.

\paragraph{Settings of the output block.}
After the final layer of message passing, we aggregate the node features in the graph through mean pooling. Then, we use a linear layer to map the $256$-dimensional graph-level features to $256$-dimensional output features, followed by a SiLU activation layer with residual connections. Next, the output is mapped to a scalar value through a linear transformation layer to complete our task.
\begin{table*}[htbp]
  \vspace{-1mm}
  \caption{Comparison between UPDDFormer, WPDDFormer, and other baselines in terms of test MAE on the JARVIS dataset. The best results are shown in \textbf{bold} and the second best results are shown with \underline{underlines}. The results reported for PotNet and Conformer in the table are those obtained from training using their published code.}
  \vspace{-1mm}
  \label{con_JARVIS}
  \centering
  \resizebox{\textwidth}{!}{  
  \begin{tabular}{lccccccccc}
    \toprule
    \multirow{2}{*}{\textbf{Method}} & \multicolumn{1}{c}{\textbf{Formation Energy}} & \multicolumn{1}{c}{\textbf{Bandgap(OPT)}} & \multicolumn{1}{c}{\textbf{Total Energy}} & \multicolumn{1}{c}{\textbf{Ehull}} & \multicolumn{1}{c}{\textbf{Bandgap(MBJ)}}  &\multicolumn{1}{c}{\textbf{Bulk(Kv)}} & \multicolumn{1}{c}{\textbf{Shear(Gv)}} & \multicolumn{1}{c}{\textbf{SLME(\%)}} & \multicolumn{1}{c}{\textbf{Spillage}}
    \\
    \cmidrule(r){2-10}
     & eV/atom &  eV &  eV/atom & eV & eV  &GPa &  GPa &No unit &  No unit 
    \\
    \midrule
    CFID (2018)& 14  & 0.30 & 240 & 220 & 0.53   &  -- &  --  &  -- &  --  \\
    CGCNN(2018) & 63  & 0.20 & 78 & 170 &  0.41 & 14.47  &  11.75 & 8.022 &  0.454    \\
    SchNet(2018) & 45  & 0.19 & 47 & 140 &  0.43 & 14.33  & 10.67 & -- &  --    \\
    MEGNET(2019) & 47  & 0.145 & 58 & 84 &  0.34   & 15.11  & 13.09 &--  & --  \\
    GATGNN(2020) & 47  & 0.17 & 56 & 120 &   0.51  & 14.32  &  12.48 & 7.504 & 0.431  \\
    ALIGNN(2021) & 33.1  & 0.142 & 37 & 76 &   0.31 &  10.40  & 9.481 & 5.145 &  0.389   \\
    M3GNet(2022) & 39.0  & 0.145 & 41 & 95 &   0.36 &  -- &  --  &  -- &  --    \\
    Matformer(2022) & 32.5  & 0.137 & 35 & 64 & 0.30& 11.21  & 10.76 & 5.260 & 0.398  \\
    PotNet(2023) & 29.4  & 0.127 & 32 &  55 &0.27 & 10.06 & 8.883& -- &  --  \\
    CrysMMNet(2023) & 28.0  & 0.128 & 34 & --  & 0.278 & 9.625  & \underline{8.471} & -- &  --  \\
    CrysDiff (2024) & 29.0  & 0.131 & 34 & 62 &  0.287 & 9.875  & 9.193&5.030 & \underline{0.358} \\
    Crystalformer(2024)  & 30.6  & 0.128 & 32 & 46 &  0.274 & -- & --& -- & -- \\
    eComFormer(2024)  & 28.4  & 0.124 & 32 & 44 &  0.28 & 10.79  & 9.826 & 4.610 &  0.373\\
    iComFormer(2024) & 27.2  & \underline{0.122} & 28.8 & 47 &  \underline{0.26}& 9.617  & 9.098 & 4.583 & 0.360\\
    
    \midrule
    \textbf{UPDDFormer}   & \underline{26.7} & \underline{0.122} & \underline{28.7} & \underline{40.6} &  \underline{0.26} &\underline{9.456} & 8.738 & \underline{4.539} & 0.364  \\
    \textbf{WPDDFormer}   & \textbf{25.7} & \textbf{0.119} & \textbf{27.6} & \textbf{35.5} &  \textbf{0.249}  & \textbf{9.224} & \textbf{8.441} & \textbf{4.441} & \textbf{0.344} \\
    \bottomrule
  \end{tabular}
  }
\end{table*}

\section{Hyperparameter Settings of PDDFormer}
\label{Hyperparameterset}
In this subsection, we present the detailed hyperparameter settings of the WPDDFormer and UPDDFormer for different tasks. We slightly tuned the parameters of our method for the material datasets, and further adjustments are expected to yield higher performance in different tasks. 

\paragraph{JARVIS.} We show the model settings of WPDDFormer in Table \ref{JARVIS_Para}. The evaluation metric for these tasks is the test mean absolute error (MAE), batch size of $64$, weight decay ~\cite{loshchilov2017decoupled} set to 1e-5. Specifically, the WPDDFormer was trained using the MAE loss function and the Adam optimizer ~\cite{kingma2014adam}. For the formation energy, total energy, band gap (OPT), and Ehull tasks, the model was trained for 400 epochs. The initial learning rate for the band gap (OPT) task was set to 0.0005, while the initial learning rate for the other tasks was set to 0.001. For the band gap (MBJ), SLME, bulk modulus (Kv), and shear modulus (Gv) tasks, the model was trained for 300 epochs with an initial learning rate of 0.001. The Spillage task was trained for 200 epochs with an initial learning rate of 0.001.

The parameter settings for UPDDFormer are the same as those for WPDDFormer across different tasks, except for the Ehull task, where the model was trained for 500 epochs.

\paragraph{The Materials Project.}
We present the model settings for WPDDFormer in Table \ref{MP_Para}. For the Materials Project dataset, all models are trained using the MAE loss function, with the Adam optimizer and Onecycle scheduler. The batch size is set to $64$, and weight decay is set to 1e-5 ~\cite{loshchilov2017decoupled}. Specifically, the formation energy model is trained for 500 epochs, the band gap model for 400 epochs, and the bulk modulus and shear modulus models for 300 epochs each, with an initial learning rate of 0.001. The settings for UPDDFormer are consistent with those of WPDDFormer.

\section{Invariance Properties}
\label{Invariance}
\subsection{Definition}
According to ~\cite{yan2022periodic}, we represent a crystal structure with a triple (X, P, L), where $(X, P) \in U$, defined as follows: $X = [x_1, \ldots, x_N] \in \mathbb{R}^{d \times N}$ represents the states of N atoms in the unit cell, $P = [p_1, \ldots, p_N] \in \mathbb{R}^{3 \times N}$ denotes the 3D Cartesian coordinates of these atoms, $L = [\ell_1, \ell_2, \ell_3] \in \mathbb{R}^{3 \times 3}$ is the lattice vector matrix. The infinite crystal structure is:
\begin{equation}
\begin{array}{@{}c@{}c@{}}
    \tilde{P} =\left \{ \tilde{p}_{i}= {p}_{i} + {h}_{1}{l}_{1} + {h}_{2}{l}_{2}+{h}_{3}{l}_{3}\right \} \\  \left \{\: {h}_{1},{h}_{2},{h}_{3} \in \mathbb{Z} ,i\in \mathbb{Z},1\le i\le n \right \} , \: \\
    \tilde{X} = \left \{ \tilde{x} _{i} = x_{i}\mid i\in \mathbb{Z} ,1\le i\le n\right \} 
\end{array}
\end{equation}

The coordinates of the n points are defined within the unit cell U as determined by L, meaning their fractional coordinates are $L^{-1}P \in [0, 1)^{3 \times N}$. When the overall network architecture is viewed as a function $f(X, P, L) \rightarrow \mathcal{X}$, they satisfy the following invariance properties.

The unique geometric prior knowledge of crystals includes two distinct physical constraints and symmetries: E(3) invariance within the unit cell and periodic invariance.

\textbf{Definition 7: Unit Cell E(3) Invariance.} Following Matformer ~\cite{yan2022periodic}, A function $\mathit{f}:\left ( \mathcal{X}, \mathcal{P},\mathcal{L}\right ) \to \mathcal {Y}'$ is unit cell $E(3)$ invariant if, for all $Q \in \mathbb{R}^{3 \times 3}$, where $|Q| = 1$, and $b \in \mathbb{R}^3$, we have $\mathit{f}\left ( \mathcal{X}, \mathcal{P},\mathcal{L} \right ) = \mathit{f}\left ( \mathcal{X}, Q\mathcal{P}+b, Q\mathcal{L} \right ) $.

In other words, the crystal structure remains unchanged when the position matrix $\mathcal{P}$ of the unit cell structure undergoes rotation, translation, or reflection.

Moreover, different minimal repeatable structures can be used to represent the same crystal. These different crystal structure representations $(X, P, L)$ introduce a constraint known as periodic invariance. Two periodic transformations can generate different minimal unit cell representations for the same crystal structure, including shifting the periodic boundary and changing the periodic pattern while maintaining the same unit cell volume.

\textbf{Definition 8: Periodic Invariance.} Following Matformer ~\cite{yan2022periodic}, A function $\mathit{f}:\left ( \mathcal{X}, \mathcal{P},\mathcal{L} \right ) \to \mathcal{Y}'$ is periodically invariant if, for any possible minimal unit cell representation $\mathcal{M}' = \left ( \mathcal{X}', \mathcal{P}', \mathcal{L}' \right )$ of a given infinite crystal structure $\left ( \bar{\mathcal{X}}, \bar{\mathcal{P}} \right )$, we have $\mathit{f}\left ( \mathcal{X}, \mathcal{P}, \mathcal{L} \right ) = \mathit{f}\left ( \mathcal{X}', \mathcal{P}', \mathcal{L}' \right )$.

\subsection{Proofs of invariance}

\paragraph{Proof of Unit Cell E(3) Invariance and Periodic Invariance.} 
If the PDD multi-edge graph we construct exhibits E(3) invariance and periodic invariance, then every step in the crystal graph construction process must conform to the crystal constraints. Therefore, we analyze the construction process of the crystal graph to progressively demonstrate E(3) invariance and periodic invariance.

First, we construct a crystal graph with $n$ nodes using a minimal unit cell structure containing $n$ atoms. This step has been handled by the JARVIS and MP datasets. Since all minimal unit cell structures for a given crystal share the same number of atoms and corresponding atomic features, this step is E(3) invariant and periodically invariant.

After determining the selection of atoms, we begin to establish edge information connecting neighboring nodes for each atom. An edge is established from node $j$ to node $i$ when the Euclidean distance $\left | e_{j'i} \right | ^{2}$ between a duplicate of $j$ and $i$ satisfies $\left | e_{j'i} \right | ^{2} = \left | p_{j} + k'_{1} l_{1} + k'_{2} l_{2} + k'_{3} l_{3} - p_{i} \right | ^{2} \le r$, 
where $r \in \mathbb{R}$ is the cutoff radius. 
We select the nearest $t$ edges within the cutoff radius, each with a corresponding edge feature $\left | e_{j'i} \right | ^{2}$. 
The Euclidean distance $\left | e_{j'i} \right | ^{2}$ between duplicates $j$ and $i$ remains invariant under E(3) transformations and different representations of the unit cell structure. Thus, the neighborhood information for node $i$ is E(3) invariant and periodically invariant.

Finally, we establish the crystal structure representations for WPDD and UPDD. For WPDD, we select $k$ nearest neighbors based on Euclidean distance to create the corresponding WPDD row for each node $i$. For UPDD, we center around node $i$ and select atoms from the reconstructed unit cell to create the corresponding UPDD row for each node $i$ based on Euclidean distance. The Euclidean distance remains invariant under E(3) transformations and different unit cell structures. Thus, the PDD row of node $i$ is both E(3) invariant and periodic invariant.

By combining these three steps in the construction process of crystal graphs, we complete the proof that the proposed PDD crystal graph representation is E(3) invariant and periodic invariant.

\begin{table}[htbp]
  \vspace{-2mm}
  \caption{Comparison of test MAE between UPDDFormer, WPDDFormer, and other baselines on the Materials Project dataset.}
  \vspace{-1mm}
  \label{Con-MP}
  \centering
  \renewcommand{\arraystretch}{1.2}  
  \resizebox{1.0\linewidth}{!}{  
  \begin{tabular}{lccccc}
    \toprule
    \multirow{2}{*}{\textbf{Method}} & \multicolumn{1}{c}{\textbf{FE}} & \multicolumn{1}{c}{\textbf{Band Gap}} & \multicolumn{1}{c}{\textbf{Bulk}} & \multicolumn{1}{c}{\textbf{Shear}} \\
    \cmidrule(r){2-5}
    & eV/atom & eV & log(GPa) & log(GPa) \\
    \midrule
    CGCNN & 31 & 0.292 & 47 & 77 \\
    SchNet & 33 & 0.345 & 66 & 99 \\
    MEGNET & 30 & 0.307 & 60 & 99 \\
    GATGNN & 33 & 0.280 & 45 & 75 \\
    ALIGNN & 22 & 0.218 & 51 & 78 \\
    M3GNet & 24 & 0.247 & 50 & 87 \\
    Matformer & 21.0 & 0.211 & 43 & 73 \\
    PotNet & 18.8 & 0.204 & 40 & 65 \\
    CrysMMNet & 20.0 & 0.197 & 38 & \underline{62} \\
    Crystalformer & 18.6 & 0.198 & 37.7 & 68.9 \\
    eComFormer & 18.16 & 0.202 & 41.7 & 72.9 \\
    iComFormer & 18.26 & 0.193 & 38.0 & 63.7 \\
    \midrule
    \textbf{UPDDFormer} & \underline{16.96} & \underline{0.189} & \underline{37.0} & 67.0 \\
    \textbf{WPDDFormer} & \textbf{16.03} & \textbf{0.187} & \textbf{32.3} & \textbf{60.5} \\
    \bottomrule
  \end{tabular}
  }
  \vspace{-2mm}
\end{table}

\begin{table*}[htbp]
  \caption{Comparison of test set MAE for WPDD with different numbers of neighbors $k$ on the JARVIS dataset. The best results are shown in \textbf{bold}.}
  \label{JARVI_k}
  \centering
  \begin{tabular}{l *{6}{c} r} 
    \toprule
     & & \multicolumn{1}{c}{Formation Energy} & \multicolumn{1}{c}{Bandgap(OPT)} & \multicolumn{1}{c}{Total Energy} & \multicolumn{1}{c}{Ehull} & \multicolumn{1}{c}{Bandgap(MBJ)} \\               
    \cmidrule(r){3-7}
    Method    & k & eV/atom &  eV &  eV/atom & eV & eV  \\
    \midrule
        {WPDDFormer}  &60 & {25.8} & {0.123} &  \textbf{26.9} &  {36.5}  &  {0.250}  \\
    \textbf{WPDDFormer}  &{92} & \textbf{25.7} & \textbf{0.119} & {27.6} & \textbf{35.5} &  {0.249}  \\ 
        {WPDDFormer} &150  & {26.3} & {0.120} & {27.8} & {37.7} &  \textbf{0.239}  \\
        \bottomrule
  \end{tabular}
\end{table*}

\begin{table*}[htbp]
  \caption{Comparison of test set MAE for WPDD with different numbers of neighbors $k$ on the Materials Project dataset.}
  \label{MP_k}
  \centering
  \begin{tabular}{l *{5}{c} r}
    \toprule
   & & \multicolumn{1}{c}{Formation Energy} & \multicolumn{1}{c}{Band Gap} & &\multicolumn{1}{c}{Bulk Moduli} & \multicolumn{1}{c}{Shear Moduli}  \\
    \cmidrule(r){3-4}\cmidrule(r){6-7}
    Method   &k  & eV/atom &  eV &  k& log(GPa) &  log(GPa)   \\
    \midrule
    \textbf{WPDDFormer}  &92 &\textbf{16.03} & {0.187} &92& \textbf{0.0323} & \textbf{0.0605}  \\
    WPDDFormer  &180 & 16.29 & {0.186} &120 & {0.0333} & {0.0619}  \\
    WPDDFormer  &300 & {16.24} & \textbf{0.184} &160 &{0.0332} & {0.0611}  \\
    \bottomrule
  \end{tabular}
\end{table*}

\begin{table*}[htbp]
  \caption{Whether to include a comparison of WPDD's test MAE on the JARVIS dataset. The best results are shown in \textbf{bold}.}
  \label{JARVI_ablation}
  \centering
  \begin{tabular}{l *{5}{c} r} 
    \toprule
     & \multicolumn{1}{c}{Formation Energy} & \multicolumn{1}{c}{Bandgap(OPT)} & \multicolumn{1}{c}{Total Energy} & \multicolumn{1}{c}{Ehull} & \multicolumn{1}{c}{Bandgap(MBJ)}  \\               
    \cmidrule(r){2-6}
    Method     & eV/atom &  eV &  eV/atom & eV & eV  \\
    \midrule
    No PDD & {27.0} & {0.122} & {28.0} & {42.6} &  {0.261}   \\
    WPDDFormer   & \textbf{25.7} & \textbf{0.119} & \textbf{27.6} & \textbf{35.5} &  \textbf{0.249}  \\ 
    \bottomrule
  \end{tabular}
\end{table*}

\begin{figure*}[htbp]
    \centering
  \begin{subfigure}{0.24\textwidth}
    \centering
    \includegraphics[width=\linewidth]{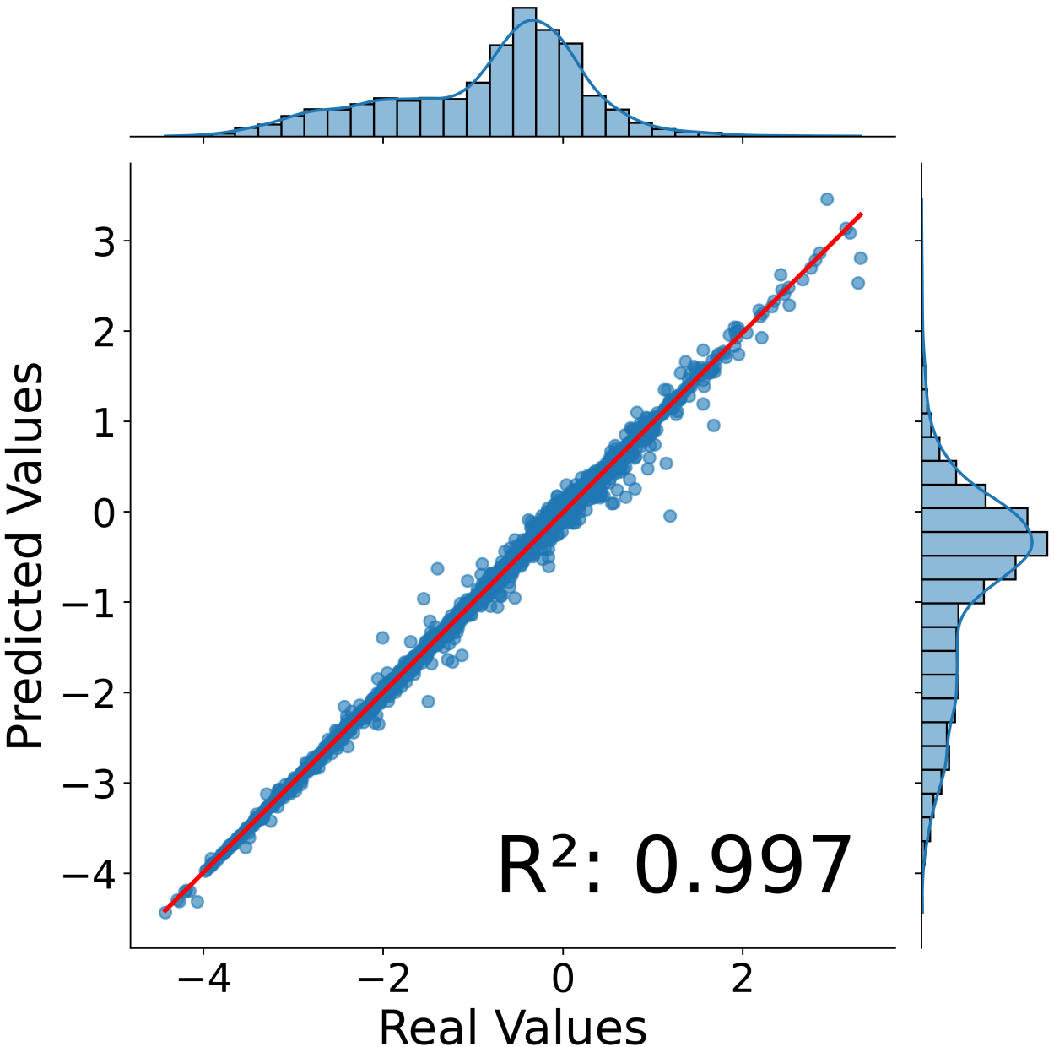}
    \caption{JARVIS-FE}
    \label{WPDD}
  \end{subfigure}
    \hspace{0.5mm}
  \begin{subfigure}{0.24\textwidth}
    \centering
    \includegraphics[width=\linewidth]{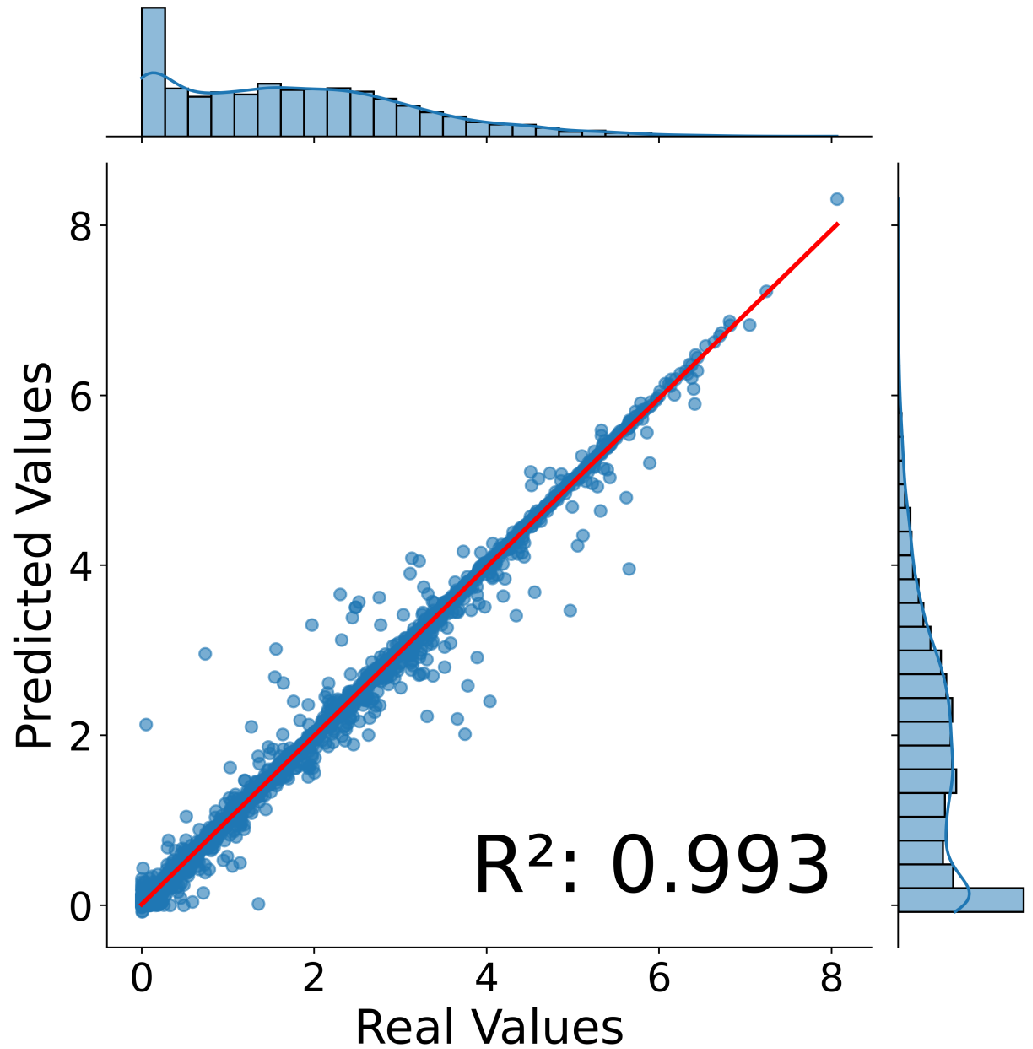}
    \caption{JARVIS-Ehull}
    \label{JARVIS-Ehull}
  \end{subfigure}
    \hspace{0.5mm}
    \begin{subfigure}{0.24\textwidth}
    \centering
    \includegraphics[width=\linewidth]{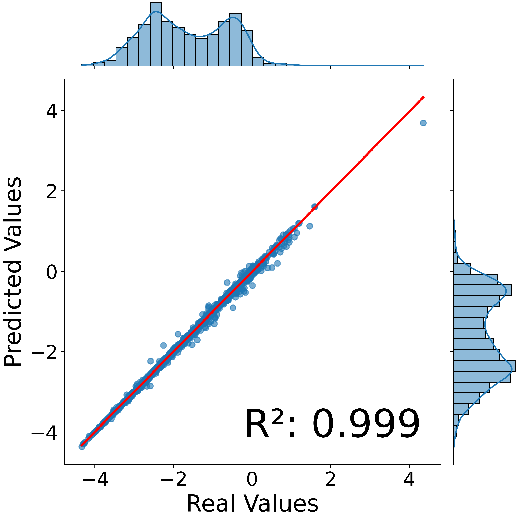}
    \caption{MP-FE}
    \label{EUPDD}
  \end{subfigure}
      \begin{subfigure}{0.24\textwidth}
    \centering
    \includegraphics[width=\linewidth]{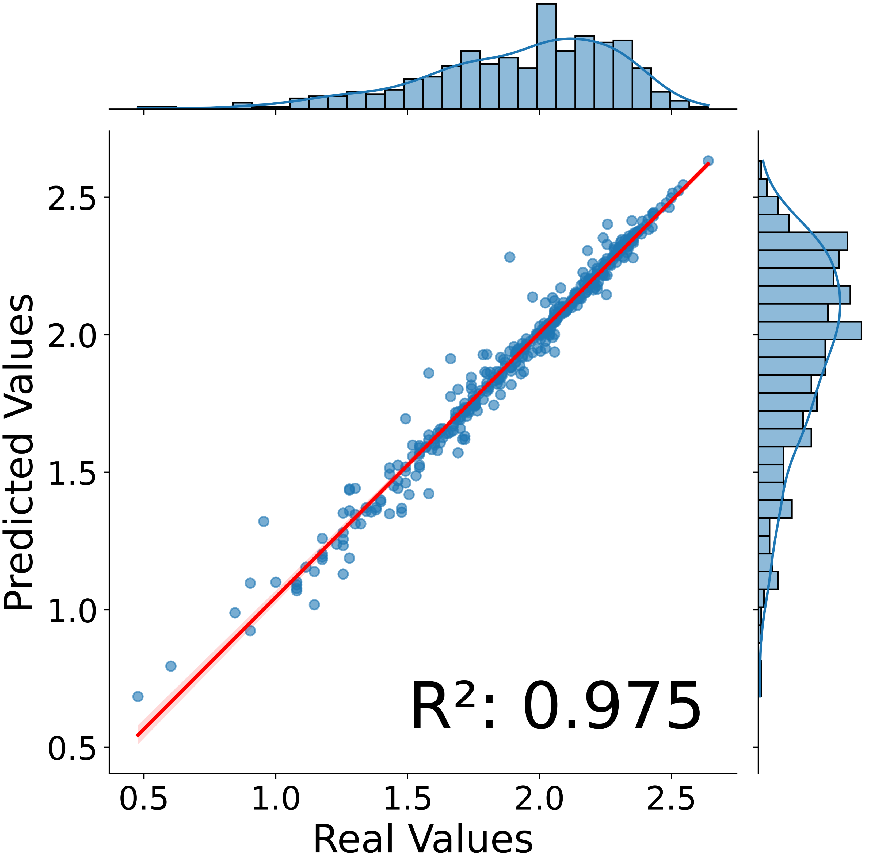}
    \caption{MP-Bulk}
    \label{EUPDD}
  \end{subfigure}
  \caption{ 
The coefficient of determination for WPDDFormer's predictions is presented. The scatter plots reflect the differences between our predicted values and the actual values, while the bar charts show the frequency of the values.}
  \label{R2}
\end{figure*}

\begin{figure*}[t]
  \centering
    \centering
    \includegraphics[width=\linewidth]{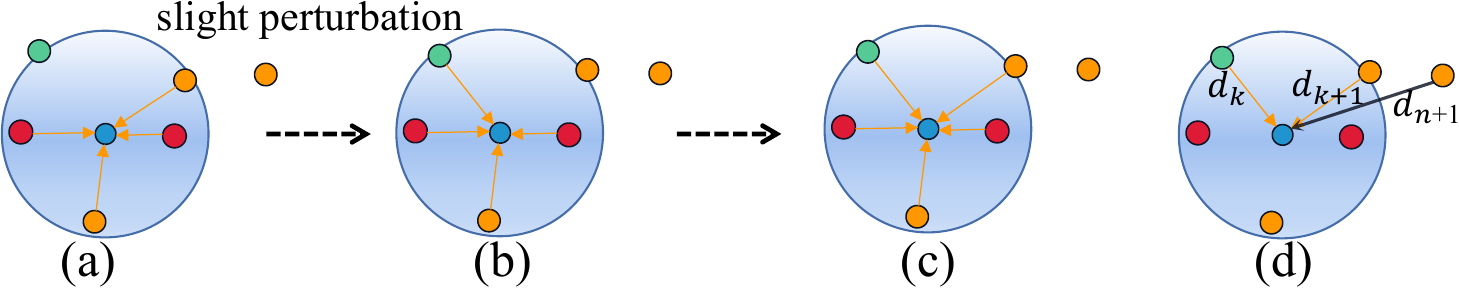}
  \caption{The different neighbor selection under slight perturbations.}
  \label{rongcha}
    \vspace{-3mm}
\end{figure*}

\section{More experimental}

\subsection{Complete experimental data.}
In this section, we provide the complete baseline model data and additional experimental results compared in the main text.

Baseline methods include CFID ~\cite{choudhary2018machine}, CGCNN ~\cite{xie2018crystal}, SchNet ~\cite{schutt2017schnet}, MEGNET ~\cite{chen2019graph}, GATGNN ~\cite{louis2020graph}, ALIGNN ~\cite{choudhary2021atomistic}, M3GNet ~\cite{chen2022universal}, Matformer ~\cite{yan2022periodic}, PotNet ~\cite{lin2023efficient}, CrysMMNet ~\cite{das2023crysmmnet}, CrysDiff ~\cite{song2024diffusion}, Crystalformer ~\cite{taniai2024crystalformer}, and ComFormer ~\cite{yan2024complete}.
For all baseline methods on the material datasets, we report the results provided in the cited papers or those reproduced based on their official codes.

\paragraph{JARVIS and MP.} As shown in Tables \ref{con_JARVIS} and \ref{Con-MP}, WPDDFormer achieved the best results across all 13 tasks on these two datasets, a performance that no previous model has achieved, highlighting the strong superiority of WPDDFormer. For the Bulk Moduli (Kv) and Shear Moduli (Gv) tasks in JARVIS, 19,680 training samples were used, while 9,066 and 11,375 training samples were used for the SLME (\%) and Spillage tasks, respectively. For the Bulk Moduli (Kv) and Shear Moduli (Gv) tasks in MP, 5,450 training samples were used. This demonstrates WPDDFormer’s adaptability to tasks with varying data scales. 
UPDDFormer surpasses WPDDFormer in inference speed while also demonstrating strong performance compared to previous models, achieving second-best results in 10 out of the 13 tasks.

We selected $R_{2}$ (coefficient of determination) as the evaluation metric. $R_{2}$
  is one of the indicators for measuring the goodness-of-fit of a model, representing the extent to which the predicted values explain the actual values. As shown in Figure \ref{R2}, we conducted experiments on the formation energy and Ehull metrics from the JARVIS dataset, as well as the formation energy and bulk moduli from the MP dataset. 
The experimental results show that the $R_{2}$ accuracy of our model has exceeded 0.99 in datasets with a larger number of training elements. Even on the Bulk property with smaller data volumes, the model performs well, demonstrating its superior performance.

\subsection{The number of neighbors ( k ) of WPDD}

In this section, we investigate the effect of the number of neighbors with different cutoff radii on the WPDD experiment. In the main text, we set the number of neighbors $k = 92$, which matches the dimensionality of the atomic feature embeddings used by CGCNN. However, in practical applications, it is necessary to determine a sufficiently large $k$ in advance to ensure completeness for any test crystal, especially in extreme cases where $k$ must be greater than the number of atoms in any test crystal ~\cite{yan2024complete}. Therefore, to ensure completeness on the JARVIS and MP datasets, we calculated the maximum number of atoms in each dataset, which is $140$ for JARVIS, $152$ for MP’s bulk and shear, and $296$ for the rest. As a result, in the experiments, the maximum number of neighbors for the JARVIS and MP datasets with different cutoff radii were selected as $150$, $160$, and $300$, respectively. Therefore, we can ensure completeness for general-purpose datasets.

From the experimental results in Tables \ref{JARVI_k} and \ref{MP_k}, it can be seen that the choice of neighbors with different cutoff radii causes fluctuations in the model's performance. However, its performance still shows the best results compared to the other models presented in the main text.

\subsection{More ablation}
To verify the effectiveness of incorporating PDD descriptors into crystal graph construction, we present in Table \ref{JARVI_ablation} the impact on experimental results on the JARVIS dataset. With minimal changes to the WPDDFormer model parameters, we investigated the effect of removing WPDD information on the experiments. 
It can be observed that after incorporating WPDD information, the model's performance significantly improved, highlighting the importance of introducing WPDD information.

\section{Continuous tolerance $\mathcal{T}$}
Given that the experimentally measured unit cell and atomic coordinates are inevitably affected by atomic vibrations and measurement noise, slight perturbations in the atomic coordinates may occur. Therefore, during the construction of the multi-edge crystal graph, when selecting the t nearest edges within a cutoff radius, the chosen neighboring nodes may change, as shown in Figure \ref{rongcha} (a) and (b). This variation in neighboring nodes $j$ leads to changes in the atomic information of the neighboring nodes, resulting in discontinuities in the construction of the multi-edge crystal graph. To eliminate the influence of atomic perturbations on the neighbor selection and ensure the continuity of the crystal graph under perturbations, we define the concept of continuous tolerance $\mathcal{T}$ to guarantee the continuity of the constructed crystal graph.

Since the distances of atomic perturbations are typically on the order of sub-angstrom (Å), specifically, in common atomic structures or crystals, slight perturbations are generally less than $\mathcal{T} < 10^{-2}$. For larger perturbations (e.g., $>\mathcal{T} = 10^{-2})$, the continuity issue may no longer be effective. Therefore, when selecting neighbors, we can set a continuous tolerance value $\mathcal{T}$ in advance. When we select the t nearest edges within the cutoff radius, if the distance of the (t+1)-th edge from node $i$ minus the distance of the t-th edge is less than the continuous tolerance, i.e., $d_{k+1} - d_k < T$, we include this neighbor in the graph construction as well. This process continues until the distance of the (n+1)-th edge from node i minus the distance of the n-th edge exceeds the continuous tolerance cutoff, i.e., $d_{n+1} - d_n > \mathcal{T}$. The n neighbors at this point are the selected nodes for constructing the crystal graph, as shown in Figure \ref{rongcha} (d), resulting in the final neighbor selection shown in Figure \ref{rongcha} (c). This approach ensures that the neighbor selection in the crystal graph construction does not change under atomic perturbations, and by using Lemma 1, we prove the continuity of the (W/U)PDD crystal graph we have constructed.

\section{Limitation}

PDD demonstrates exceptional discriminative capability: with the number of neighbors k=100, it can distinguish all known crystal structures in the Cambridge Structural Database (CSD). However, a theoretical scenario exists where, with an infinitely large unit cell, no matter the value of k, there will always be a larger unit cell that renders k insufficient to ensure completeness. Therefore, absolute completeness cannot be achieved in a theoretical sense. Nonetheless, such extreme scenarios exist only in theoretical derivations and are exceedingly improbable, even in theory. More importantly, these situations do not occur in the real world. Thus, our approach satisfies the general completeness requirements for almost all materials under perturbations.

\end{document}